\documentclass[%
  superscriptaddress
  ,preprint
  ,amsmath,amssymb
  ,aps
  ,showkeys
  ,nofootinbib
]{revtex4-1}

\usepackage[]{graphicx}

\begin{document}

\title{Open-ended Evolution and a Mechanism of Novelties in Web Services}

\author{Takashi Ikegami}
\email{ikeg@sacral.c.u-tokyo.ac.jp}
\affiliation{Department of General Systems Studies, The University of Tokyo, Komaba, Tokyo 153-8902, Japan}

\author{Yasuhiro Hashimoto}
\email{hashi@sacral.c.u-tokyo.ac.jp}
\affiliation{Department of General Systems Studies, The University of Tokyo, Komaba, Tokyo 153-8902, Japan}

\author{Mizuki Oka}
\email{mizuki@cs.tsukuba.ac.jp}
\affiliation{Department of Computer Science, University of Tsukuba, Ibaraki, Japan}

\begin{abstract}
  Analogous to living ecosystems in nature, web services form an artificial ecosystem consisting of many tags and their associated media, such as photographs, movies, and web pages created by human users. Concerning biological ecosystems, we regard tag as a species and human as a hidden environmental resource. We subsequently analyze the evolution of the web services, in particular social tagging systems, with respect to the self-organization of new tags. The evolution of new combinations of tags is analyzed as the open-ended evolution (OEE) index. The tag meaning is computed by the types of associated tags; tags that vary their meanings temporally exist. We argue that such tags are the examples of OEE.%
\end{abstract}


\maketitle

\section{Introduction}
Open-ended evolution (OEE) is a characteristic of nonbiological yet evolutionary processes that can be observed primarily in human technologies such as computers, airplanes, smart phones, and other advanced technologies that support our daily  life. We herein define OEE as the progressive improvement in quality and invention of novelty in these technologies. The development of personal computers over the decades is a good example. 
OEE is not only caused by a singular development of technologies or ideas behind each product, but also the network effect of various technologies and ideas. For example, the development of transistors will lead to better smart phones that will change the methods of human communication. Changes in human communication will determine the direction of smart phone development, and eventually require new transistor designs.
This type of recursive processes has been used in the development of human technologies and human society for a long time. OEE occurs in these systems even without the Darwinian processes. 

Different focal points exist in OEE and different definitions and hallmarks have been discussed in the Alife community~\cite{Taylor2016}. Similar to Banzah and Beson, we think that the most important aspect of OEE is the novelty production. While increasing the degrees of freedom, a system will produce novelty rather than exhibiting the entropic growth, i.e. randomly generated ``mutants'' merely dominate a whole system. Erwin further discusses in the same paper~\cite{Taylor2016} that novelty and innovations are different. Innovation requires ecological or biological success, while novelty does not. 

Mark Bedau et al.~\cite{Bedau2011} investigated the evolution of US patents and observed fruitful insights on OEE. One of them is the ultrahigh evolutionary speed of patents observed in their analysis. It may not be merely explained by a simple 
idea of evolution in ecosystems such as point mutation and/or natural selection. Such accelerated evolution dynamics are expected to be found in biological systems. 

We herein focus on the emergence of OEE in the evolution of web services. The most typical characteristic of web services is that they contain human activities in their loop, i.e., the preferences of human users change the evolutionary dynamics of the web service, and changes in the web service changes the preferences. The recursive evolution between human and the service is the primary mechanism of OEE in the web services. That is, we discuss the third mechanism of OEE, not in human technologies or in genetic systems, but in  the human machine (web) interfaces.

The purpose of this paper is to present examples of OEE in the web service through analysis and to measure OEE events from empirical data. Further, we attempt to re-interpret the findings in the analysis in terms of biological evolution.

Regarding social tagging systems (STSs) (in this study, we choose Delicious, Flickr, and RoomClip), we focus on the evolution of new tags and their new combinations. STSs reflect how our cognitive space is expanded and facilitated by the creation of new words (tags). New words provide us with new concepts and emotions in daily life, and the creation of new action patterns triggered by new ideas and emotions can result in the creation of many new ideas~\cite{Kauffman2000}. 
That is, OEE in social tagging dynamics is a co-evolutionary system between human behaviors and vocabulary, wherein new tags, which are introduced by cultural preference, generate new behavioral patterns in human activities. Such a co-evolutionary process recurrently changes the vocabulary of the tags.

The mathematical treatment of word creation and selection in social tagging analyses is often studied by a simple Yule--Simon (YS) process~\cite{Cattuto2006,Koya2019}: reproduction of tags and introduction of new tags through random mutations. However, some studies have assumed a correlation between novelties~\cite{Tria2014} or a latent semantic structure behind word occurrences~\cite{Cattuto2009,Bedau2011}. To observe patterns and structures in web evolution, we define and analyze the combinatorial novelty rather than the novelty of simple new tag creations. Specifically, we investigate the novelties that result from combining tags as opposed to the random introduction of a new tag. Ackley argues that OEE requires indefinite scalability. We propose that novelty introduced by the combinatorial space of tags overcomes the non-scalable property of a system, i.e., even with a finite number of tags, their combination diversity can grow exponentially. 

We report the transitions in the semantics of tag usages in STS web services. Tag usages in the early stages may become different as the time elapses. We found such examples and successfully visualized them by revealing the semantics of the tag usages. Changes in semantics in the usages of tags are a primary discovery of this study. This is an evidence of OEE in the STS services. The co-occurring tag sets of a single submission will change depending on the social or users' context of tag usages. Many tags will either be fixed or exhibit punctuated equilibrium (used for one purpose and subsequently transition abruptly to another purpose), while some tags will be drifting forever. In terms of genetic systems, this drift can be considered as sequential pre-adaptation, i.e., the function of a phenotype is not fixed and will change over time. We will focus on this new type of OEE observed in web services.

\section{Data and Analysis}

\subsection{Web data preparation}
Different web services exhibit different dynamics but share many statistical properties in common: Heaps laws, Zipf’s law, etc. We analyze tagged data obtained from three different web services: Delicious (social bookmarking service for storing, sharing, and discovering web bookmarks), Flickr (photograph-sharing service), and RoomClip (interior photograph-sharing service). The data consist of a list of annotations, where each annotation contains the time stamp of when it was created, ID of the photograph or website, ID of the user who posted the photograph or website, and the tag string. The data from Delicious are from almost four years since the inception of the service, and the total numbers of distinct words and annotations are approximately $2.5\times 10^6$ and $1.4\times 10^8$, respectively. The data from Flickr are from almost two years since the inception of the service, and the total numbers of distinct words and annotations are approximately $1.6\times 10^6$ and $1.1\times 10^8$, respectively. Further, the data from RoomClip are from almost four years since the inception of the service, and the total numbers of distinct words and annotations are approximately $3.3\times 10^5$ and $8.8\times 10^6$, respectively. Over $7\times 10^4$ users posted at least one photograph. The Delicious and Flickr data are from \cite{Gorlitz2008}, and the RoomClip data were provided directly by the service provider. It is noteworthy that the data from RoomClip are much smaller compared to those from Delicious and Flickr.

Additionally, Flickr and RoomClip are called narrow tagging and Delicious is called broad tagging; this is because in Delicious, users can tag the posted bookmark but in the other two services, tagging is restricted to the submitted users. Additionally, RoomClip and Flickr exhibit the social network nature, i.e., they are used to communicate among users and follow-follower networks exist.

\subsection{Data Analysis}
Because each post contains many tags in general for all three STSs, it is worth computing the similarity between the co-occurring tags in submissions between different weeks. Using the dataset from all the three STS services, we calculated the following quantities:

\begin{enumerate}
\item The statistics over the entire time period: the number of distinct tags, the number of new tags, the production rate per day, and a set of the co-occurring tags in the same submission.
\item Similarity between the tag usages:  
  The vocabulary similarity (distribution similarity) of each tag between the given two weeks is calculated using the Jansen--Shannon divergence (JSD):
  \begin{align}
    d^k_\text{JS}(p_t||p_s)&=\left[d_\text{KL}(p^k_t||q^k)+d_\text{KL}(p^k_s||q^k)\right]/2,\nonumber\\
    q^k&=(p^k_t+p^k_s)/2,\nonumber
  \end{align}
  where $p_t$ and $p_s$ are the probability distributions of co-occurring tags in the same post over a week $t$ and $s$, respectively, with respect to a tag $k$. The term $d_\text{KL}(p||q)$ expresses the Kullback--Leibler divergence of the probability distribution $q$ to $p$. (The value of $d_\text{JS}$ is in the range between 0 and $\log_a2$;
  here, $a=2$ and therefore $0\le d_\text{JS}\le 1$.)
\item Similarity between human users' preferences: A similar method to the similarity between the tag usages is applied to measure the similarity between human users. All the tags posted by each human user are used to detect users that are similar to each other on the tag usage profiles. 
  A smaller $d_\text{JS}$ value means that the two distributions are similar to each other. We interpret here that  those users $t$ and $s$ exhibit a similar nature.
\end{enumerate}

\subsection{Yule--Simon Model}
Yule introduced the model to describe the evolution of biological species and the genus. Species in a genus can generate a new species and sometimes even create a new genus. This process is described by two types of mutations: a new species or a new genus is created with certain probabilities. Later, by reorganizing the formulation, it is formally described as the YS stochastic process~\cite{Yule1925,Simon1955} as follows: 

The YS model generates a series of symbols $\{\sigma_1, \sigma_2,$\\
$\sigma_3,\ldots, \sigma_t\}$, where $\sigma_t$ denotes a tag drawn from a pool of tags with a probability $1-\alpha$. A new tag is introduced with a probability $\alpha$. The original model can be solved analytically and the characteristic properties of such processes can be obtained exactly as in Zipf's law.

A new YS model replaces a tag with a set of tags, and generates the time series of the set of tags. At each time step, a set of tags is added sequentially, where a new type of tag is stochastically introduced in the same way as the original YS model. In this model, the novelty creation does not necessarily require the introduction of new tags; a novel combinatorial usage of the existing tags can also create a pairwise novelty.

Similarly, with the original YS model, a pair of tags will be uniformly and randomly chosen among the existing tags in the time series, which is equivalent to the process referred to as the preferential attachment effect~\cite{Newman2003}. 
In the following, we will use the new YS model as a reference for interpreting the combinatorial usage of tags in the empirical data. Although it is a over-simplified model, we will show through analysis that the empirical data cannot be described with a random selection hypothesis assumed in the new YS model.

\section{Results of the Data Analysis}

\subsection{Time evolution of the number of tags and posts}
From our previous studies, we found that the volume of the entire dictionary (of tags) increases typically as $t^{\beta}$, by denoting the number of annotations as $t$, which is known as Heaps' law. The value of the exponent $\beta$ is between 0.7 and 1 empirically.

As shown in Figure \ref{fig:IncreaseInPosts}, a soft transition of the cumulative number of posts occurs as a function of service age. 
This transition is recognized as the shift from the initial to the mature stage of the service, in which the increase in the number of users, their activities, the number of posts, and the diversity such as tag vocabulary are accelerated. It could be caused by system changes in the web service. In fact, management companies often introduce a search engine optimization (SEO) at certain timings. It could also correspond to the critical point obtained from the analysis using Hawkes process conducted on ``likes'' event sequences on photographs~\cite{Ikegami2017}. The results in \cite{Ikegami2017} indicated that the service evolved to a near (sub-) critical point. This observation implies that the service is maintained at a critical point to become sensitive to the reaction from the users.

In general, quantity changes to quality above a certain system size. Such examples can be found in natural and artificial systems. In the case of artificial systems, a simple boid system (swarm model) shows qualitatively different dynamics beyond a critical flock size (approximately 10,000)~\cite{Ikegami2017}. In the human community size, a critical community size can be assumed (e.g., Dunbar's number) owing to the bounded cognitive capability of agents and the resulting stability of the community~\cite{Dunbar1992}. The current example provides yet another example of such size-dependent system responses.

\subsection{Time evolution of Novelties}
\begin{figure*}[t]
  \centering
  \includegraphics[width=\linewidth]{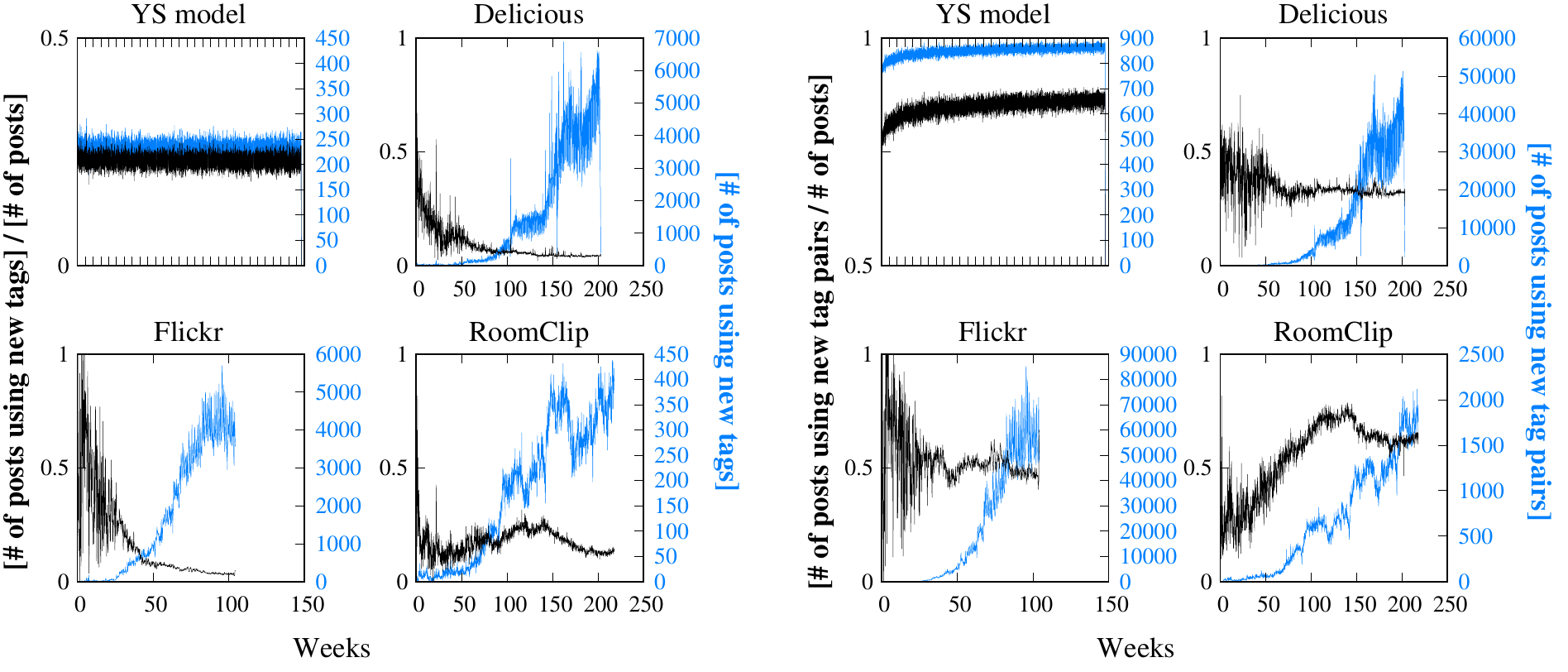}\\
  \caption{Single tag novelty rate per day (left) and a pairwise novelty rate per day (right) are computed among three different services and the Yule--Simon (YS) model. The horizontal axis is the number of weeks since the inception of each service (In the YS model, a physical time unit does not exist). Black and blue lines are, respectively, the proportion of daily posts using new tags against the total number of daily posts and the raw number of daily posts using new tags.%
  }
  \label{fig:SingleAndPairwiseNoveltyRate}
\end{figure*}

Because our analysis is on the tagging data, we regard a novel tag as a novelty characteristic in each web service. We compared different web services with respect to their novelty rates, which are defined as the creation rates of new tags per day. Unlike the most Alife systems, those web services continue producing new tags and the number of distinct tags increase continually; meanwhile, the novelty rate decreases as the user population size increases, as shown in Figure \ref{fig:SingleAndPairwiseNoveltyRate}.

\begin{figure*}[t]
  \centering
  \includegraphics[width=\linewidth]{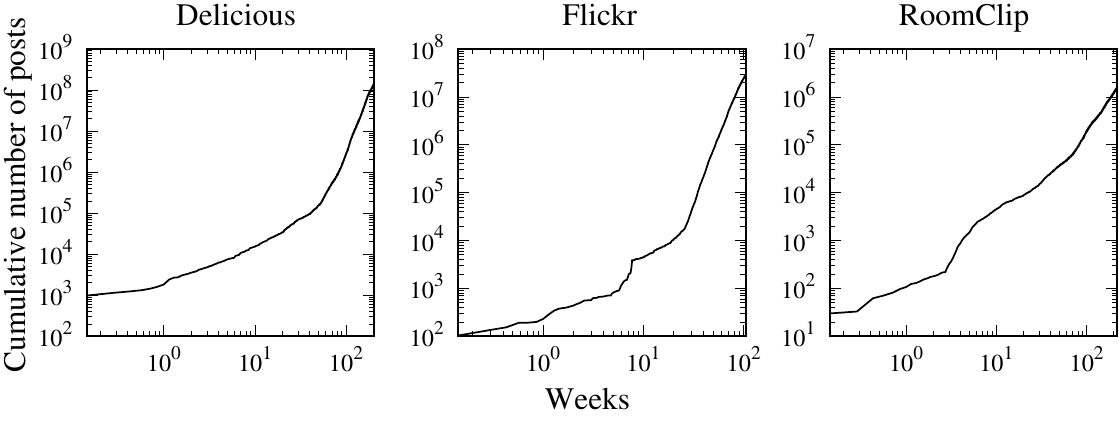}\\
  \caption{%
    Cumulative number of posts as a function of the number of weeks since the inception of each service.%
  }
  \label{fig:IncreaseInPosts}
\end{figure*}

Figure \ref{fig:SingleAndPairwiseNoveltyRate} (left) shows the novelty rate of new tags for different web services computed from the temporal production of tags over time. The number of posts that contain novel tags (blue) increases as the whole number of posts increases,  as shown in Figure \ref{fig:IncreaseInPosts}. Meanwhile, the proportion of posts using new tags (black) decreases to 10\% eventually, or below of the whole posts per day for each web service. For RoomClip, the peak of the proportion in approximately 100 to 150 weeks is consistent with the inflection point in Figure \ref{fig:IncreaseInPosts}.
The three web services share two common features: they ``grow'' temporally, and the novelty production rate for tags decreases eventually.

The YS model was used as a reference to interpret the empirical data.
It assumes neither structure in a pool of tags nor new tags. Further, the mutation rate is fixed in YS, and it is sustained as a constant in the figure.

\subsection{Evolution of the Pairwise Novelties}
\begin{figure*}[t]
  \centering
  \includegraphics[width=\linewidth]{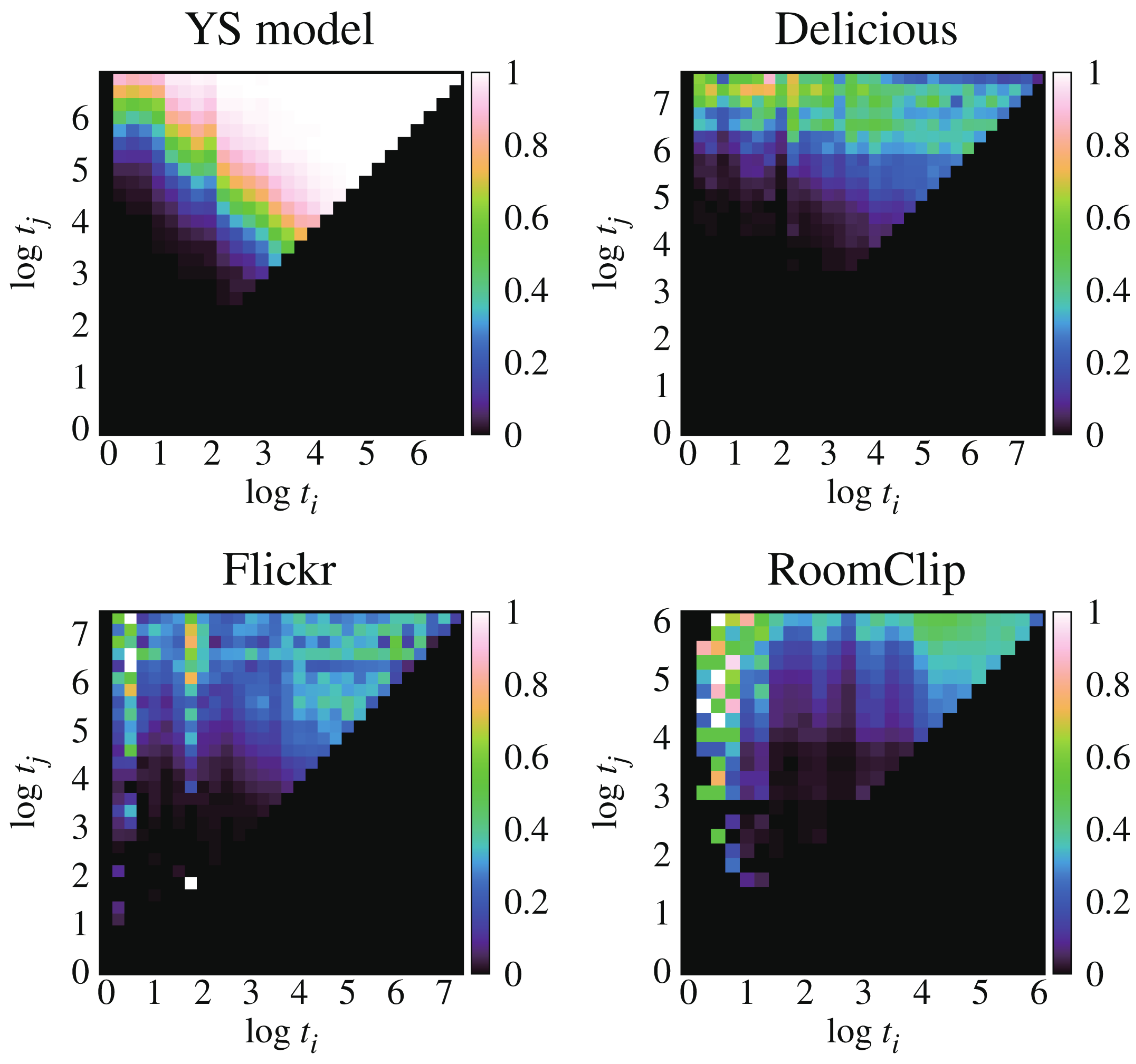}\\
  \caption{Probability of two tags, born at the time signified by y-axis and x-axis, respectively, co-used in the same post for the first time among all co-usages at that time. The brighter color indicates a higher probability. Compared with the reference YS model, tag usages in the web services are biased toward the upper right corner (see the interpretation in the text).}
  \label{fig:PairWiseNovelty}
\end{figure*}

In addition to creating new tags, there is another mechanism that introduces novelty in the system.
The second mechanism is the development of a new combinatorial usage of tags.
Figure \ref{fig:SingleAndPairwiseNoveltyRate} (right) is drawn by counting the distinct pairs used for each submission. The pairwise novelty was also computed from the modified YS model and the computed novelty rate was overlaid in the figure. Unlike the single-tag novelty rate, it increases slightly for the YS model. The empirical data exhibit the gradual decline of these web services. This discrepancy is explained by the fact that the YS model assumed random pairing from the word pool, while the empirical data used biased pairs of tags. In fact, we visualized the creation frequency of a new combination of two existing tags (born at time signified by the y-axis and x-axis, respectively) being co-used in one submission (Figure \ref{fig:PairWiseNovelty}). The brighter color indicates the higher probability of obtaining such new pairs. The figure implies that tag usages in the web services are biased toward the upper right corner, i.e., newly generated tags are likely to be used in the recent submissions. It is noteworthy that those three services are different in tagging manners, but the tendencies shown in Figure \ref{fig:PairWiseNovelty} are rather universal for STS services. 

\subsection{Meaning shift}
\begin{figure*}[t]
  \centering
  \includegraphics[width=\linewidth]{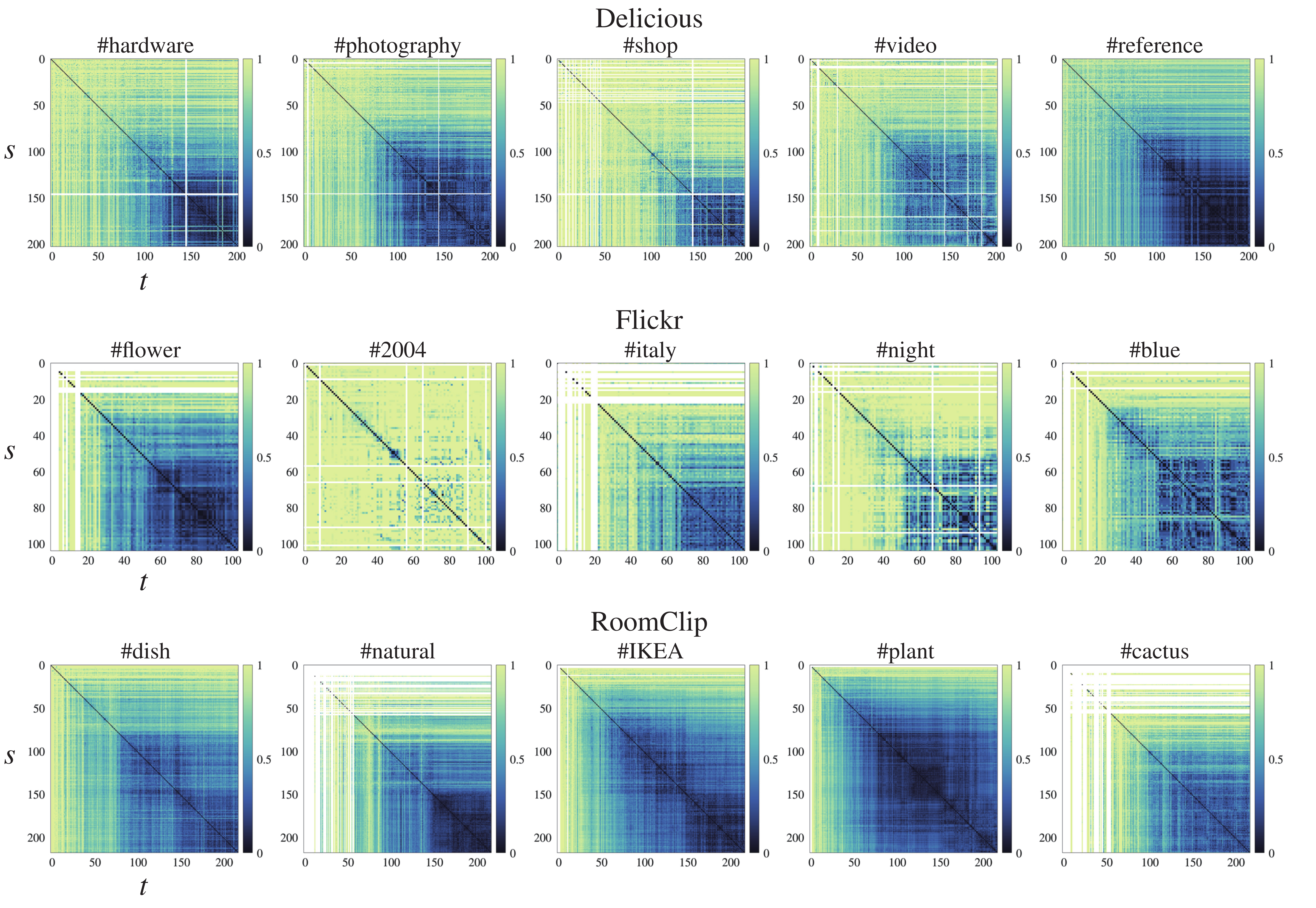}\\
  \caption{Partial example of semantic shift in tags in each service. Meaning of tag $k$ at time $t$ is given by the function $f_k(t)$, which is the distribution of co-existing tags associated with a post on the $t$-th week (we omit co-existing tags whose contribution is smaller than 1\% of the total co-occurrences for the tag of focus). We used the Jansen--Shannon divergence (JSD) to compute the difference in meaning between $f_k(t)$ and $f_k(s)$. The darker color indicates a smaller difference. The accumulation of dark color at the bottom right corner means that the meaning of the tag $k$ is becoming fixed.}
  \label{fig:JSDMatrix}
\end{figure*}

The meaning of a tag can be identified as a set of concurrently used tags within the same submissions. 
In Figure \ref{fig:PairWiseNovelty}, it is difficult to ascertain how tag meanings change over time. Thus, in Figure \ref{fig:JSDMatrix}, we computed the distribution function $f_k(t)$ of concurrently used tags with a tag $k$ over a week at time $t$. We subsequently computed the JSD between two functions, $f_k(t)$ and $f_k(s)$  for every tag $k$. If the two functions are identical, then the JSD becomes zero. If the two functions are vastly different, then the JSD will be close to 1, and we say that the same tag has different meanings if the JSD deviates from 0. 
In Figure \ref{fig:JSDMatrix}, all three services are depicted and the darker color indicates that the JSD approaches to zero.
From the observation, we can constract a graph separately for two typical cases: one with transitions among different meanings and one that converges at the end. Tags are either ones i) with a gradually darker color at the bottom right corner, i.e., converging to the common meaning over submissions, or  ii) with no dark colors in the entire process, i.e., no common meanings emerge. The meaning shifts from one entity to another.

\begin{figure*}[t]
  \centering
  \includegraphics[width=\linewidth]{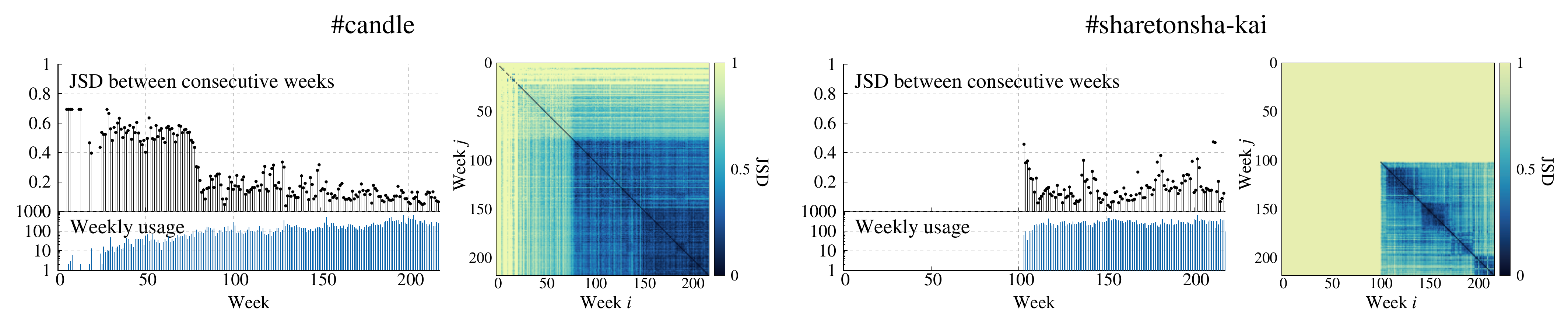}\\
  \caption{Examples of converging (left; \#candle) and non-converging (right; \#sharetonsha-kai) cases; ``\#sharentosha-kai'' means ``fashionable/cool''+``club''. In both cases, the upper left panel shows the transition of the JSD between two consecutive weeks, which exhibits a decreasing behavior and some spikes in the left and right case, respectively.}
  \label{fig:consecutiveJSD}
\end{figure*}

Most tags for all three services tend to approache a stable meaning. 
The tag ``\#cactus'' of RoomClip is a typical example of the case i) . Its usage converges gradually to a common value. On the contrary, tag ``\#2004'' in Flickr corresponds to the case ii). 
In between i) and ii), an interesting exception exists. Figure.\ref{fig:consecutiveJSD} (right) displays the time evolution of the JSD of ``wandering'' tags. A mysterious tag ``\#sharetonsha-kai'' from RoomClip exists; it is a slang in the Kyushu Island of Japan that means  ``fashionable/cool''+``club.''  This tag became a buzz word and has spread from Kyushu Island to cover the entire western Japan district. It became a popular tag that can be associated with many other tags. Further, this tag reflects the background user community. This example shows a wandering behavior and is captured by the spiky trains of the JSD between two consecutive weeks.

\subsection{Community Similarity}
The other changes underlying the evolution of the web service are the changes in the user community structure.
We applied a clustering method to the user community to identify the community structure emerging in the RoomClip service. The setup and procedure of the analysis are as follows:
\begin{enumerate}
\item Extract the users whose number of posts during the data period are greater than or equal to 100. This threshold is determined ad hoc to ensure that their vocabulary size is sufficiently large.
\item Calculate a probability distribution of used words for all extracted users.
\item Calculate similarity between the probability distributions for every pair of the extracted users, and define a user similarity network in word usage.
\item Traversing from a loosely to densely connected network by changing a similarity threshold, verify the connectivity of the highly productive users.
\end{enumerate}

Herein, we define the novelty production rate of the user as the total number of words that were created by a user and used by more than 100 other users. The vocabulary similarity between a pair of users was evaluated using the JSD as in the previous sections.

\begin{figure*}[t]
  \centering
  \includegraphics[width=\linewidth]{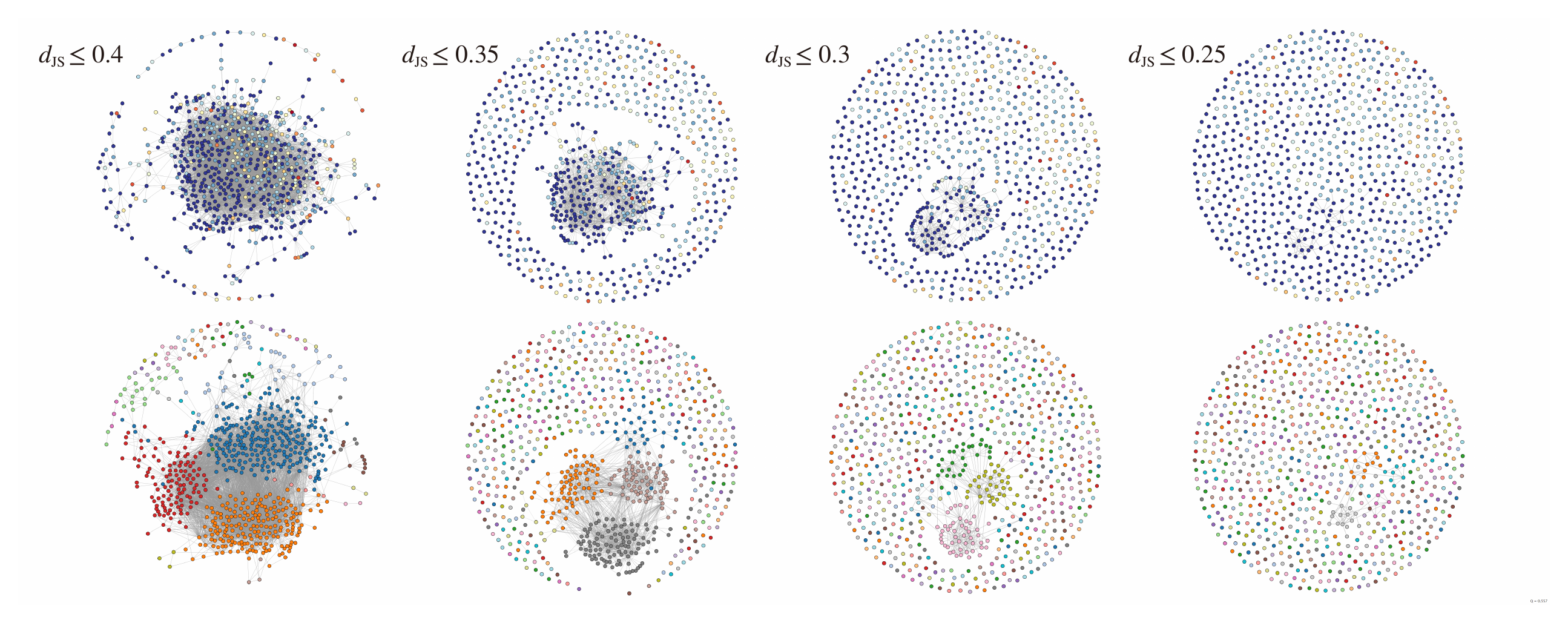}\\
  \caption{Tag novelty rate (the upper column) and the community structure (the lower column) of the user similarity network on RoomClip. Each node is a user, and is connected if $d_{\text{JS}}$ is less than the threshold values 0.4, 0.35, 0.3, and 0.25 from left to right. The top four figures show the number of tags created by individuals in color; shifting from blue to yellow to red means that they have created more words. The bottom four figures show the community structures detected in each network. In these figures, the isolated users (in the peripheral) create novel tags at a high rate.}
  \label{fig:user_similarity_network}
\end{figure*}

The result of the community comparison is shown in Figure \ref{fig:user_similarity_network}. The laterally aligned top-four figures show networks with different threshold values of $d_\text{JS}$ that decrease from left to right. The far-left case ($d_\text{JS}\le 0.4$) exhibits a typical core-periphery structure~\cite{Borgatti1999} that contains a densely connected part (core) surrounded by loosely connected parts (periphery). Using a smaller threshold value means that a more strongly-connected part is focused, where a subnetwork associated with the core part of the core-periphery structure is obtained. A community structure means that a loosely connected set of densely connected subgraphs arises with such small threshold values. This point becomes clearer when we apply a community detection method to the networks. The bottom-four figures exhibit the community structures detected by the modularity optimization method~\cite{Newman2006} \footnote{Extracting densely connected subnetworks by comparing with a randomly connected network.}. The obtained modularity values (0.35, 0.42, 0.49, and 0.56 from left to right) indicate that the community structure becomes relatively salient in a strongly connected part of the network. In fact, three self-organized entities share the similar word-usage profiles.

Meanwhile, the colors of the nodes in the top-four figures show the novelty production rate of each user (i.e., red color corresponds to high and blue color to low novelty rates. In the core part, the novelty production rate of users exhibit a relatively lower value (see the case of $d_\text{JS}\le 0.3$ and $0.25$). 
Users who create new words that are also used by a certain number of other users are located outside of such community structures (i.e., a peripheral parts of the network). We found in the previous studies that the novelty creation rate increases in the higher-order cliques in the core part.
In other words, users in the peripheral parts and higher-order cliques have higher new tag production rates~\cite{Ikegami2017}. 

\section{Discussions}
The mechanism of OEE in a web service is based on the interplay among the active users, user community, and service. We proposed and visualized new mechanisms of the OEE by analyzing three different web services. Possible scenarios of web evolution in terms of tags are as follows: i) Population size increases by changing the system meta-parameters, ii) The creation of new tags will be enhanced when the population size increases beyond a critical value, iii) A set of tags that specifies a common usage of a tag (semantics) shifts spontaneously. The semantics appear 
to converge as time elapses for most of the tags; however, some exceptions exist. The semantics of a tag will wander without converging, iv) A user community develops sub communities of users with similar profiles (i.e., sharing similar tag usages).
Point iii) is related to OEE. Namely, some popular tags will exhibit OEE with the sufficient set of other tags. We hypothesize that the evolution of the community structure iv) can enhance/accelerate the evolution of novelties that in turn results in a self-organized community structure. Therefore, the co-evolution nature of the human user and the web-serviced interface can be established. Hence, a possible OEE mechanism from this study involves two interacting layers: social tagging system and the human community. 

Although this scenario should be refined with further analysis, we expect that it corresponds to Ackley's definition of the indefinite scalability of OEE~\cite{Ackley2014}. That is, ``supporting open-ended computational growth without requiring substantial re-engineering.'' The growth and expansion in the novelty searching space by combinatorial effects results in a potential ``door-opening'' innovation.

In addition, as discussed in \cite{Taylor2016}, the present work provides another example of adaptive novelty, because the evolution of a web service can utilize the exponentially large numbers of new combinations of existing fixed-meaning tags that will generate qualitatively new niches for the underlying user community. However, we have not successfully measured the effects on the advantages in the community.

Compared with biological evolution, we discovered that 
a set of tags can be a genotype and the associated photograph can be a phenotype. Therefore, the characteristic of a phenotype is given by a combination of tags, and the meaning shift discussed herein is interpreted as ``speciation,'' i.e., emergence of a new species. If a series of new speciation continues, we regard it is an evidence of OEE. Therefore, to discuss the OEE of an object, the latter must be composed of many elements. In other words, combinatorial novelty can potentially drive OEE. This idea is also related to the GARD model~\cite{Lancet2000} that uses a compositional rather than sequential genome system for transferring information、as a set of tag for each submission is not concerned with the order. OEE is thus expected in the GARD model from this aspect.

\section*{Acknowledgements}
This research was supported by MEXT as ``Challenging Research on Post-K Computer''—Modeling and Application of Multiple Interaction of Social and Economic Phenomena (hp160264). It is also partially supported by Grand-in-Aids for Scientific Research (B) ``Analysis of Internet services as Biological Evolution systems'' (17H01821). We also acknowledge RoomClip Inc.~for providing their RoomClip data for analysis.

\bibliographystyle{apa-good}
\bibliography{ikegami}

\clearpage

\clearpage

\clearpage

\clearpage

\clearpage

\clearpage

\end{document}